# Dynamic DNA Processing: A Microcode Model of Cell Differentiation

Barry D. Jacobson

*Abstract*—**A general theoretical framework is put forth to organize and understand various observed phenomena and mathematical relationships in the field of molecular biology. By modeling each cell in eukaryotic organisms as a processor having a unique set of allowed states, represented by a specific DNA sequence, we demonstrate a method by which gene expression can be regulated. As the theory is developed, we suggest reasons for the complementary, quaternary (4-base) coding scheme used in most eukaryotes. A role for transposable elements is suggested, as is a role for the abundance of noncoding DNA, along with a clearly-defined method by which single nucleotide polymorphisms (SNP's) may alter gene expression. The effect of various errors is considered. Finally, a mechanism for inter-processor communication is proposed to explain cell-cell recognition processes, which leads to an elucidation of a possible pathway by which nonmutagenic carcinogenic agents may act.**

*Index Terms*—**Amino acids, Bioinformatics, Biological information theory, Biomedical signal processing, Biomedical computing, Biomedical engineering, Biophysics, Cancer, Computational molecular biology, DNA, DNA computing, Gene therapy, Genetic expression, Genetic programming, Genetics, Genomics, Molecular biology, Molecular communication, Nanotechnology, Oncology, Proteins, Proteomics, Single Nucleotide Polymorphisms, Systems biology.**

## I. PREFACE

In 1997, this article was submitted to Nature after a number of years of work, but was rejected for lack of evidence after an in-depth review. At that time, sequencing of the human genome was 5 years away. After it was successfully completed for the first time in 2003 [1], the time and cost involved were still too prohibitive to make it a routine lab procedure. Because the model to be presented here depends on the existence of small polymorphisms in DNA sequences among different cell types, there was no way to verify these at the time. In the past 16 years, however, improvements in technology have decreased the time for sequencing a whole genome to the order of weeks or less, and the cost is continually coming down, as well.

In addition to this, other pieces of corroborating evidence have come to light during these 16 years that were anticipated in the original paper. An example is the recent discovery by the Project ENCODE Consortium [2] that the huge amounts of noncoding DNA (>98%) [3] may in fact not be "junk DNA" as it was commonly termed, but may serve as a set of switches to control and regulate gene expression [4]. But while the ENCODE paper relies on inferential and indirect evidence to reach this conclusion [5], it does not provide any specific mechanism by which this regulation could be achieved. Our original paper of 1997 described an orderly mathematical theory by which non-coding DNA may serve as software in a clearly-defined processing scheme. It provided a possible explanation of how even a single base-pair change, as in a single nucleotide polymorphism (SNP), may completely redirect the control logic of a cell. In addition, it suggested an important role for DNA looping, which in a recent discussion of the ENCODE work, was described as "poorly understood" in the following excerpt [6]:

> "Beyond the linear organization of genes and transcripts on chromosomes lies a more complex (and still poorly understood) network of chromosome loops and twists through which promoters and more distal elements, such as enhancers, can communicate their regulatory information to each other."

Beginning in the next section, we present the entire paper as it was first submitted, except for an additional figure and occasional minor edit, so that it might be judged on the basis of what has transpired in the intervening 16 years. Following this, in the Postscript section, we first recapitulate with a short summary of the essential points of the model, and then proceed to list what we believe are the major strengths of our approach, including several additional points that were either not known at the time or did not occur to us, that might also be understood on the basis of the original model. We also append Table 1, comparing conventional thought with our Microcode Model for easy reference.

## II. INTRODUCTION

MUCH HAS APPEARED in the literature regarding identification of the factors responsible for cell fate determination, and the mechanisms by which cells in eukaryotic organisms differentiate into their respective types which differ so vastly in form and function [7].

Along with these two issues are related questions concerning the mechanisms that govern cell-cell recognition and cell-autonomous determination, since clearly any proposed scheme to explain differentiation must account for observed phenomena of both types. Experiments in limb regeneration have shown that there exist intercalatory processes that depend on a cell's being able to recognize the

This work was supported in part by the National Institute of Health under Training Grant 5T32 DC00038.

B. D. Jacobson is with the Harvard-MIT Division of Health Sciences and Technology, Cambridge, MA 02139 USA (e-mail: bdj@alum.mit.edu).

exact positional value of its neighbor, such that intermediate values between host and graft are correctly interpolated, thus facilitating a smooth transition in tissue types between the two boundary values [8]. On the other hand, experiments have likewise demonstrated that independent of its home environment, a cell remembers its identity even when transplanted to a foreign environment [9, 10]. We might additionally seek to know where this line between heredity and environment lies for cells. Up until what point can they adapt to new surroundings, and at what point must they resign themselves to being locked into a preprogrammed fate?

Regarding the first set of questions, as to how cells with identical genetic material can differ in form and function, recent approaches have tended to focus on the existence of promoter and inhibitor regions of the genome which act as anchors for general transcription factors that serve as controllers to switch on and off the genes responsible for coding the various proteins that distinguish one cell type from another [11]. In a cell of type A, the genes manufacturing type X protein are turned on, whereas the genes manufacturing type Y protein are turned off; in cells of type B, the converse occurs. However, the obvious question then is how these regions are differentially activated in the two types of cells? What guarantees that type A cells will obtain or manufacture just those proteins that bind to type X promoters and to type Y inhibitors, and that type B cells will acquire an exact opposite set? This edifice forestalls the original question, but doesn't fully explain the prime cause. In order to logically extrapolate back to the fertilized egg of origin, it is necessary to presuppose additional layers of complexity, e.g., polarization of the parent cell such that when cell division occurs, products are unevenly distributed, thus forming a conceptual binary tree structure [12].

While the existence of promoters and inhibitors is an indisputable fact, we may stand to benefit by looking still further for additional directors of the development process. A key incentive is that in the absence of any other unifying factors, we have almost no hope of fully comprehending or intervening in the process should it be necessary—the complexity is far beyond our current capability [13]. Consider what has to happen. Suppose we are given the sequence of a key fate-determining protein product which has properly found its way into only one particular cell of a generation. From this sequence we must now calculate its structure, a difficult task. From that we hope to predict its activity. Since this is a fate-determining protein, it will bind (or cause others to bind) to at most a limited subset of DNA sequences. We must then somehow find a binding site or sites likely to be favored over all others in the entire genome, and furthermore predict what effect that will have on the transcription process. We still need a way to foretell how the newly manufactured products which were regulated by our protein will distribute themselves among the two daughter cells in the next generation. We have been forced to go to great lengths in which mistakes are easily made at every step, just to analyze a single cell in a single generation. Imagine the scenario if we tried to project 5 or 10 generations down the road.

Drawing upon various observations recorded in the literature, and where necessary, postulating the existence of certain agencies which have not been verified, the author would like to outline a theory which may partially address several of these issues as well as a number of others, some of which he has not seen raised previously, but which nevertheless may hold some significance. The specific mechanism by which this scheme can be implemented is left open, although one possible route will be illustrated.

## III. DEVELOPMENT OF MICROCODE MODEL

### A. Proposed Existence of DNA Clock

The main concept upon which this framework is based is that there should exist somewhere within the genetic material of every cell a region of DNA which serves as a clock sequence that keeps track of the cell's individual identity. (There is a large literature on the subject of biological clocks [14], however, the particular variation to be presented here does not appear to be found elsewhere.) Let us assume that in each succeeding generation, one base-pair is added onto the clock sequence as a cell undergoes replication. One could thereby distinguish on the basis of length alone, the "age" in generations of a particular cell. Now, let us further assume that in each cell of a particular generation, this sequence will differ, even very slightly, from that of all other cells in the same generation. One could then use this sequence to uniquely identify any cell in the organism, and could furthermore obtain its complete history by merely reading the clock, as we will show.

### B. Definition of Microcode

The motivation for this type of scheme comes from the microcode concept used to run the central processing units (CPUs) of many modern computers. Microcode is a series of hard-coded instructions for controlling each phase of a single processing or memory cycle. In a typical CPU, a program instruction must be fetched, decoded, and then acted upon appropriately. But to fetch an instruction from memory requires that first the address of the current instruction be looked up (it is normally stored in a special internal register of the CPU known variously as the program counter or instruction pointer). The address must then be sent over the address lines (known collectively as the address bus) towards the memory module in order to select the desired memory location from all the others. Certain control signals (strobes) are used to initiate and verify correct transmission of the address. Next, other control signals are used to initiate and verify the memory read process whereby the contents of the desired location are transmitted over the data lines (known collectively as the data bus), and latched into the CPU. At each stage, all other devices sharing the address and data buses must be locked out from transmitting any information until that bus is free. This, too, requires still other control signals (gating signals) to be transmitted at particular times to various tri-state gates which are switches that grant or block access by a device to a signal line. After the instruction is fetched, it is then examined to see if it requires any operands. For example, if the instruction is to perform an addition, the numbers to be added may possibly be in memory, or in a local storage register, or directly included in the instruction word itself. If the instruction indicates that the addends must themselves be fetched from memory, then a similar set of memory operations will be necessary for each. Next, the CPU will need to activate the appropriate logic to execute the desired mathematical or logical operation. A separate sequence of steps will also be required to store the output of the operation, in this case, the sum. It may need to be placed in a specific memory location, or in a particular register, for example.

The method used to coordinate this complicated sequence of events is that a counter triggered by a series of timed pulses continuously cycles through a predefined binary sequence. At each individual step in the cycle, the value of the counter is input to control logic which, based upon the value of the counter, generates the appropriate control and gating signals in order to open or close each line as appropriate. The control logic is called microcode to distinguish it from the regular program code (commonly known as software) which sits at a higher level of abstraction, and is not directly interfaced to the hardware. The microcode on the other hand, is usually permanently etched into a built-in read only memory unit (ROM), and physically controls the underlying hardware needed to execute the program code. What is important for our purposes is that the value of the counter itself is the input signal to the control logic. It is not being used merely to count or time the occurrence of external events, but rather, each time the counter assumes a new value, that itself is a primary event which will, after translation by the microcode, be interpreted as a command to open or close certain address and data paths, or generate particular control signals. The counter may be thought of as a rotating cylinder such as found in a child's mechanical music box on which a pattern of small teeth project that strike correspondingly aligned musical arms, thereby sounding the proper notes at the proper times.

The general term for this type of device is a finite state machine whereby the device can assume any one of a finite set of states at each of which a certain output may be generated. The value of the current state, and the inputs to the device determine what will be the next state. (This formulation is known as a Moore machine; others can be shown to be functionally equivalent [15].)

### C. Generation of DNA based Clock Sequence

Getting back to cells, for the clock sequences to exhibit variation in cells of the same generation, there must be a way to produce an asymmetrical result in the transmission of genetic material from a parent to its daughter cells such that the two descendants end up differing in one base-pair.

A possible manner in which this can occur is that in each replication cycle the cell will exploit the asymmetrical nature of the two daughter strands of DNA that are formed as a result of the unzipping of the double-stranded parent into two complementary strands. (Whereas each strand carries the same information, they are clearly not identical). Let's assume, for simplicity, that the clock sequence resides somewhere on chromosome *n*. In normal (non-clock region) replication or translation, the incoming nucleotides are strictly ordered and regulated by the existence of a complementary template to which it must base-pair. No differences in the daughters' DNA are possible from that in the parent's DNA (barring errors), since each daughter strand, although initially differing from its sibling (being its complement), will eventually bind with complementary nucleotides leaving each with the same two strands as in the parent. However, in the clock region, we have assumed that an additional base-pair is *added* onto the clock sequence of the parent, and we make the further assumption that this occurs before the final synthesis step wherein a complementary partner becomes bound to each nucleotide. In this region, no previous template exists; it is uncharted territory. Space is left for one to postulate that in the absence of a template on the opposite strand (which has since separated), selection of the incoming nucleotide is left to the last nucleotide on the *same* strand—it alone will decide who its new neighbor should be. I.e., the final nucleotide of the parent's clock sequence, will determine according to some predefined system of rules, which of the four possible bases is to be the next one added, and that will then become the final nucleotide of the daughter's clock sequence. While this may appear far-fetched, non-templated addition of single nucleotides by DNA polymerases has actually been documented as a nuisance in PCR work (albeit on one strand); and selection indeed depends on the terminal nucleotide (and the polymerase) [16, 17]. Additional justification for proposing such a scheme can be made by noting that it has been demonstrated experimentally that the overall stability of a molecule of DNA depends on interactions between nearest neighbors, with certain combinations of neighboring bases contributing more to the free energy than other combinations [18].

Let us imagine the simplest possible rule for determining the addition of new bases onto a clock sequence—a simple cyclical scheme (Fig. 1):

```
A  ->  C
C  ->  G
G  ->  T
T  ->  A
```

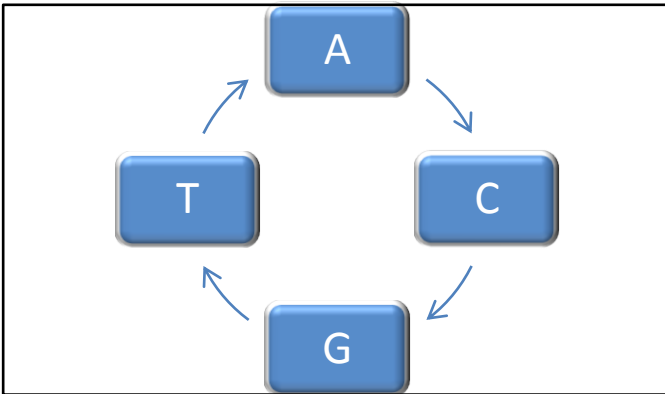

Figure 1. Simple cyclical selection rule for addition of next nucleotide in a 4-base system. Terminal nucleotide of strand determines nucleotide to be added.

If the final nucleotide of the parent's clock is an *A*, then the next nucleotide added will be a *C*, and that will become the final nucleotide of the daughter's clock, with the *A* now finding itself the next to last nucleotide in the daughter's clock. If the final nucleotide is a *C*, then the next one to be added will be a *G*, and so on, according to our rule; noting that if the final nucleotide is a *T*, then the next one will be an *A*, which will restart the cycle. Nothing useful has happened yet, until we make the following observation: Using the same cyclical scheme on the other daughter strand which is as of yet similar to its sibling, only complementary to it, will yield a *completely different* nucleotide, not the same nucleotide added to the other daughter's clock, and not even the complement of that newly added nucleotide.

To see this, consider the following example of a sequence of base-pairs:

```
X-X-X-A
| | | |
X-X-X-T
```

The double stranded DNA will unzip during replication so that we have the following two single strands:

```
X-X-X-A
| | | |

| | | |
X-X-X-T
```

If now we add an additional base onto the ends of each strand according to our cyclical selection rule, we end up with the following:

```
X-X-X-A-C
| | | | |

| | | | |
X-X-X-T-A
```

since an *A* begets a *C*, whereas a *T* begets an *A*.

Next, the two strands each take on a complementary strand in the synthesis stage to form two double stranded molecules:

```
X-X-X-A-C
| | | | |
X-X-X-T-G

X-X-X-A-T
| | | | |
X-X-X-T-A
```

One can see that the two new molecules differ in the last position.

If each of these two daughter molecules replicates, in turn, we end up with the following:

```
X-X-X-A-C-G
| | | | | |
X-X-X-T-G-C

X-X-X-A-C-A
| | | | | |
X-X-X-T-G-T

X-X-X-A-T-A
| | | | | |
X-X-X-T-A-T

X-X-X-A-T-G
| | | | | |
X-X-X-T-A-C
```

Each of the four members of this generation also has a unique clock sequence. We can continue indefinitely, and using our rule, will always be left with unduplicated sequences for all progeny in the division cycle. The author has developed a computer simulation to propagate up to 10 generations of cell divisions and has verified that this is indeed true, although it should be obvious, and can probably be proved by induction. Fig. 6 shows the clock sequences of the first 12 cells from the 10[th] generation (out of 1024 total output sequences, for space considerations). For reasons we discuss later, the corresponding amino acid sequences of the three possible reading frames for each strand are shown to the right. The source code is available from the author.

### D. Why Nature Chose a Complementary, 4-Base System

As an aside, one can now readily understand the need for quaternary (4-base) coding of genetic information. Conceivably, replication and translation machinery could just as well have been designed using a binary system alone, i.e., 2 bases. (Note that one could argue that codons would then have to be five base pairs long [$2^5 > 20$] as opposed to three [$4^3 > 20$] in order to uniquely specify the twenty amino acids [plus a few start and stop signals], resulting in decreased efficiency; however, there would be less redundancy in terms of multiple codons coding for the same amino acid, as we would then have only 32 codons, total, rather than 64. That is a significant savings, since only half as many types of nitrogenous bases, tRNA and associated enzyme machinery would be necessary for polypeptide synthesis. Conceivably, this could actually produce an overall decrease in total genomic DNA, since the enzymes and tRNA molecules are themselves complex compounds requiring a certain amount of overhead in their manufacture. Hence, it's arguable that increased efficiency would actually result.) However, in binary, this additional property of sequence uniqueness would not be possible. The reason is that suppose the only allowed bases were *A* and *T*. We might therefore have a sequence of DNA in the clock region similar to the following:

```
X-X-X-A
| | | |
X-X-X-T
```

When that cell divides, it will have to obey the following selection rule shown in Fig. 2 (no other possibilities):

```
A --> T
T --> A
```

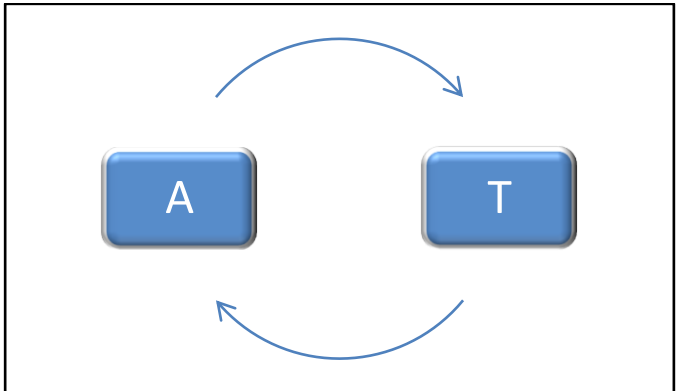

Figure 2. Simple modification of cyclical selection rule will not preserve sequence uniqueness in a 2-base system.

After unzipping, the two strands will look like this:

```
X-X-X-A
| | | |

| | | |
X-X-X-T
```

Next, the strands would each add an additional base according to the selection rule (*A* adds *T*, and *T* adds *A*):

```
X-X-X-A-T
| | | | |

| | | | |
X-X-X-T-A
```

Finally, after the synthesis step, we end up with:

```
X-X-X-A-T
| | | | |
X-X-X-T-A

X-X-X-A-T
| | | | |
X-X-X-T-A
```

The result would be that the identical clock sequence would appear in both cells. That would violate the requirement that each cell of a generation have a unique clock. Such a scheme, therefore, will not work in binary. Only a more complex selection rule that adds two new bases at a time to each strand (not to be confused with a base pair), the identities and order of which would depend on the final two bases of that strand, will preserve clock sequence uniqueness in binary. An example (which the reader should verify) would be as in Fig. 3:

```
AA --> AT
AT --> TA
TA --> TT
TT --> AA
```

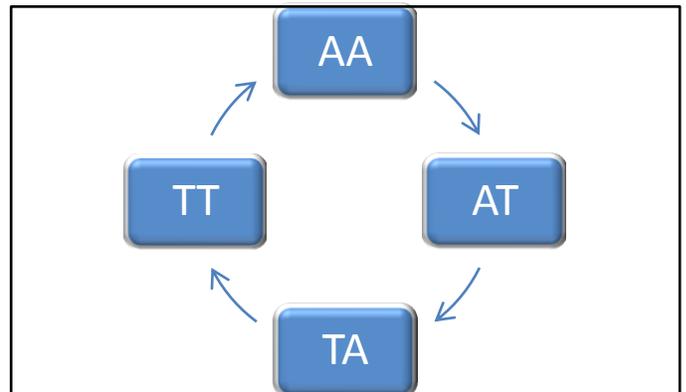

Figure 3. More complex cyclical selection rule which will preserve sequence uniqueness in a 2-base system. Final two nucleotides of strand determine next two to be added.

However, DNA polymerases could never make a living if they worked in that manner, since most of their business comes from painstaking, single-base, templated synthesis (would be true in binary, as well). In addition, as noted earlier, it is likely that the overall thermodynamic stability of a molecule of DNA depends most heavily on interactions between nearest neighbors; not on interactions between neighboring bases more distantly located.

Similarly, the need for complementary base-pairing should now be apparent. Previously, it has been suggested that the reason the DNA molecule in higher organisms is composed of two strands is the ability that confers to perform error-correction [19]. If a base becomes altered, the odds are that its partner will not be affected simultaneously. The mismatch can therefore be corrected by appropriate measures, restoring the original sequence. However, that alone does not explain the need for complementary coding. (Although, clearly, the hydrogen bonds between corresponding bases that hold the two strands together are best aligned in the case of Watson-Crick complements; we are speaking on a conceptual level, i.e., does this fact bestow any special advantage upon the cell.) Suppose that $A$ always paired with $A$, $C$ with $C$, $G$ with $G$, and $T$ with $T$. There would then be an even simpler error correction method—just locate base pairs that don't have an identical partner on the opposite strand. According to our scheme, however, if bases were identically rather than complementarily paired, there would be no asymmetry to give rise to daughter strands that differ in one position.

Note that this additive mechanism with which we have been working is not the only way to exploit the asymmetry between the complementary strands. It has been noted that the bases themselves are not very different from each other chemically, and at times, certain bases transform by means of chemical reactions such as deamination or depurination into another base [20]. Commonly, these are thought of as errors which must be fixed. However, maybe they could also serve a purpose by allowing a certain systematic variability at certain points to implement a changing clock sequence. Since the strands are complementary, the changes that occur on one strand would not be expected to correlate with the changes on the opposite strand. Still another way to make the argument plausible is to recall that the replication fork is inherently asymmetrical in the 5'-to-3' direction as compared to the 3'-to-5' direction [21]. The author has chosen to focus on the additive approach, however, as it is the easiest to conceptualize.

*E. Minimum Necessary Clock Sequence Length*

One point worth calculating is the minimum length $L$ of a clock sequence that would be sufficient to uniquely tag every cell. Ostensibly, that would be $L = log_4 D$ where $D$ is the total number of cell divisions in the life of an organism. In humans $D$ has been estimated as $10^{16}$. The length $L$ would then be 27. Certainly, that is not a very long sequence compared with the total 3 x $10^9$ base-pairs in the human genome.

*F. Regulation of Gene Expression*

The next question is how the cell can utilize this clock sequence to regulate gene expression. A possible route would seem to be the various looping, transposition, recombination and dislocation events that have been observed for many years in genetic material [22-24]. Consider a hypothetical chromosome containing the following sequence located somewhere, which we will call a target sequence, and which is present in all cells of the organism since it is a non-clock sequence:

```
X-X-X-A-C-G-X-X-X
| | | | | | | | |
X-X-X-T-G-C-X-X-X
```

One of the four second-generation daughter cells that we examined earlier has a clock sequence that matches this sequence. It, therefore, has the ability to interact with this target sequence. This interaction may take the form of a looping, transposition, recombination or dislocation event that is initiated by that match. However, none of the other cells of that generation will have that ability, since they do not match completely. The end result of this selective interaction may be that certain genes are expressed in certain cells but not in others. One can readily imagine many reasons for this. If the DNA in a given cell is looped in a particular configuration, that may either assist or block certain transcription factors from binding to the required promoter or inhibitor regions. This could cause a differential expression of particular genes in that cell as compared to another cell which has a linear topology (Fig. 4). It is interesting to note that it is thought that the very method by which action-at-a-distance can occur between promoter regions distantly separated (in terms of numbers of intervening base-pairs) from their corresponding coding regions is just that—by means of a looping process [25].

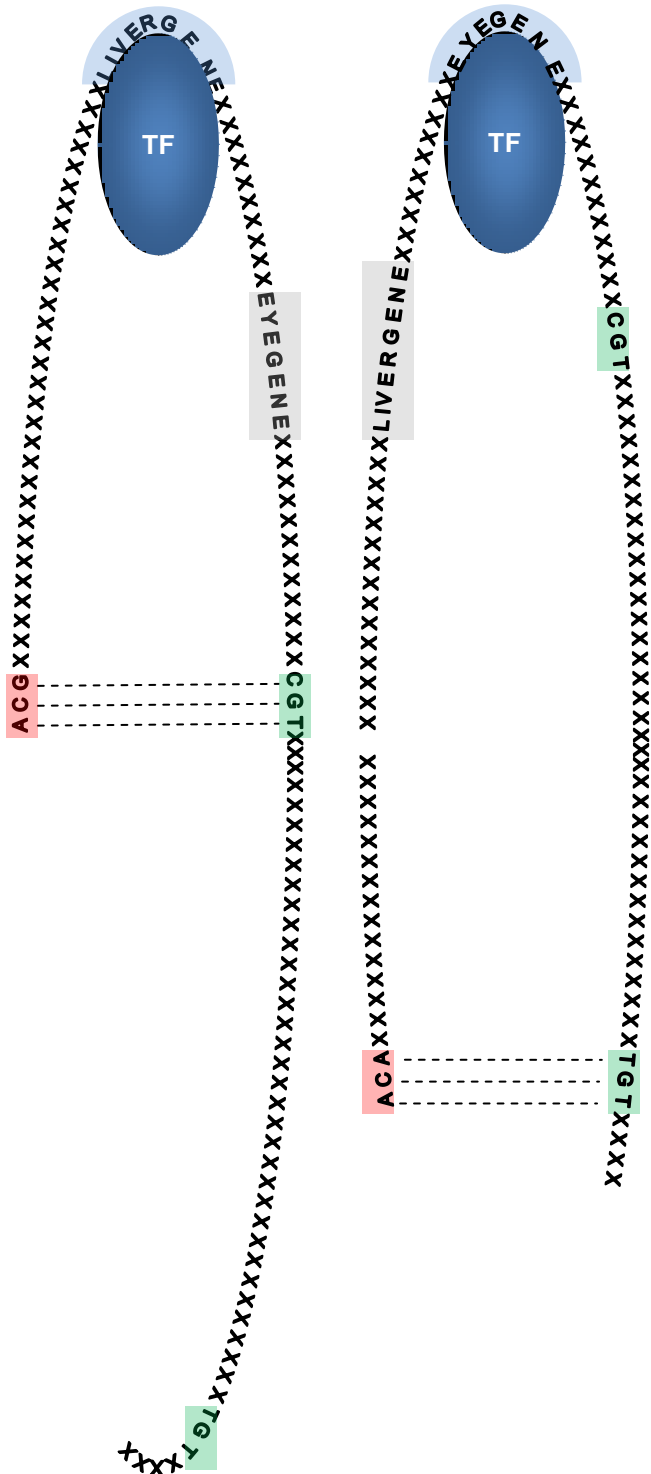

Figure 4. An example of how a single nucleotide polymorphism (SNP) in noncoding DNA can change expression of a gene. Clock sequences are shown in red; targets in green. Left panel shows cell with clock sequence *ACG*. It hybridizes with target *CGT*, but not with target *TGT*. The resultant loop topology forms an active site for a transcription factor TF (blue) at the site of a liver gene. This TF is assumed to require the vertex of a loop, whereas the eye gene has a straight topology. Right panel shows a different cell with clock sequence *ACA*. It hybridizes with target *TGT*, but not with target *CGT*. The resultant topology creates a loop vertex at the site of the eye gene, while the liver gene now has a straight topology. Thus, eye gene is now transcribed by TF, but not liver gene. (For targets, we illustrate with reverse complementary sequence for clarity.) Note that we could just as well have drawn the situation reversed, that the TF only fits in a straight topology, and is too bulky to fit in a loop vertex. We drew it this way for simplicity.

## G. Non-coding DNA and Transposable Elements

This could conceivably be a function for the abundant quantities of noncoding DNA which exist for long stretches between the coding regions [26-28]. Investigators have wondered about the repeating short patterns which are often found in these regions [29, 30]; possibly, they could serve as targets for selective matching with particular types of clock patterns.

Although we grouped together various types of interactive events (looping, transposition, recombination and dislocation), conceivably one could imagine different functions for each. We have already discussed a possibility for looping. (It will be convenient to refer to the latter three categories as relocations.) As for transposition, it is known that periodically, short sequences of DNA (transposons) move about from place to place in the genome [31]. Their exact function (if there is any beneficial one) is unknown [32]. According to our scheme, they could indeed serve a very useful purpose. We have assumed that the clock sequence resides on chromosome *n*. There must be some method by which the other chromosomes are also made aware of the current state of the clock. This could conveniently be performed by the transposons. Every so often, when the clock has reached some notable milestone, a specific transposon might find a match on the particular chromosome for which that milestone is relevant, and then proceed to attach itself, thereby informing and updating that chromosome as to the occurrence of this clock milestone. Similarly, at particular stages, selective recombination and dislocation events could occur, thereby altering the relative positioning of certain crucial chromosomal landmarks from cell to cell. Over the course of the organism's developmental history, these collective events might ultimately manifest themselves in differential gene expression. It should be mentioned that, although above we assumed for simplicity in the minimum clock length calculation that one would require a unique sequence to tag each cell, conceivably, after any relocation event, the cell could start counting again from an earlier point in the sequence (permitting sequence reuse), since the cell has already distinguished itself on the basis of its unique configuration, and is now on a different developmental path. (This actually hinges on the issue of reversibility, whether reversing a clock will reverse all the subsequent events that occurred, since. Relevant for later discussion.)

In short, a path of strategically laid out targets along the genome, will direct particular cells along particular routes, with each cell's clock acting as a private key, unlocking only those gates for which it has privileged access.

## H. Effect of Clock Errors on Development

Let us now consider the effects of certain errors on the development of an organism. If a mutation should occur in the clock which conceptually advances it, then certain stages in development will occur prematurely, skipping over necessary prerequisites. For example, there is a developmental disorder which causes the hands to grow from the shoulder. If the clock has advanced in error, instead of first producing the forearm, the organism might be prompted to initiate development of the hand. Similarly, if the clock has been set back, it will cause a delay or omission in subsequent development along that path, or possibly, duplication of a previous stage.

### I. Cancer

Next, consider what would be the result of a mutation which causes a mismatch with the next scheduled relocation milestone (a discrepancy between the clock and target). The cell will begin to drift off course. At each division, it will find no target with which to match. As a result, there will be no control mechanism to guide the cell along its proper developmental path. This may be a step in understanding cancer, where developmental braking becomes lost, and the cell divides uncontrollably. A way to see this is that in our scheme the targets serve as checkpoints, and direct each cell along a particular route where its activities will be suitably regulated. As long as it follows its proper course, its function can be specified, and quotas and limits can be placed on its various output products. That can be achieved by any of several well-known types of feedback mechanisms available to a cell for adjusting the quantity of a particular substance produced. This will be true for those factors that control cell division, as well. However, if the cell has missed a scheduled target, the feedback loop is effectively broken, since it depends on a delicate mix of counterbalancing factors being present at the proper times, and the program that the cell is to follow regarding how many divisions it is to undergo is destroyed.

Another way to conceptualize this failure is simply as an infinite loop, where the cell has no exit point since the loop test condition can never be fulfilled. I.e., in pseudocode (C programming language syntax) we have:

```
for ( clock=start_value; clock != target;
    update_clock() )
{
    ...
    perform_cell_division();
}
```

This loop can only be terminated if at some point the clock is able to match the target. If a mutation occurs which prevents this from happening, the cell will divide *ad infinitum*.

One can also see how the amplifications and deletions of various genes often seen in cancer may arise. The failure to match with a scheduled target or the coincidental formation of a match with some non-intended target due to a clock or target error may cause a conformational change in DNA which ultimately produces these effects. Abnormal translocation of chromosomal material (the significance of which is now well established) could similarly be initiated as a result of such an error. Sometimes such a translocation combines parts of two separate genes together and the result is a chimeric (fusion) protein. This is known to occur in chronic myelogenous leukemia. In other instances, an existing gene (proto-oncogene) is moved closer to a promoter region, thereby activating the oncogene. A condition in which this occurs is

Burkitt's lymphoma [33]. One of the major difficulties in understanding cancer is in sorting out which are the causes and which are the effects. The conventional view would probably differ with our model in its interpretation of these events. The normative view would likely hold that the defining moment in the initiation of tumorigenesis is the production of some abnormal (either in quantity or in character) oncogenic or chimeric protein. The exact functions of these products are currently unknown in many cases, but are widely believed to be involved in some direct or indirect way in controlling cell division or in regulating transcription of other (downstream) genes [34]. According to our model, however, the abnormal chromosomal conformation due to loss of communication between clock and target may itself be a culprit, and would not necessarily require production of the product of the immediately affected gene located at or near the breakpoint [35], although the long range result might still be the incorrect activation or inactivation of some distant gene. In general, a mismatch could come about as a result of any of the suspected causes of cancer, including genetic defects, retroviruses, or, as we will discuss later, chemical carcinogens; hence this standpoint is not incompatible with current thinking.

Thinking back to our state machine analogy, such an occurrence would correspond to a forbidden state, one in which no logic has been implemented to steer the machine from that state to another state, due to the designer's mistaken belief (or hope) that it would never be entered in the first place, since it is not normally reachable from any other state. In digital design this is a common pitfall for which engineers must constantly be on the lookout. On power-up or at some other time when a glitch occurs, a system which is supposed to cycle through what seems to be a simple, orderly set of states may lock up because the now random voltages that appear on the various flip-flops may represent a state which is not part of the sequence for which the device was intended. Provision must always be made for the device to go from any possible state to some permitted state.

### J. Cell-Cell Recognition Processes

Up until this point we have concerned ourselves solely with cell-autonomous determination, which is generally defined as those mechanisms which internally guide the cell along its proper developmental path without regard to the cell's external environment. However, these mechanisms alone are insufficient to explain differentiation, determination and development. Clearly, a mechanism exists by which the external environment, including neighboring cells can play a role in the developmental path of a given cell. There is much evidence for this [36, 37], including the grafting experiments we mentioned at the beginning of this article whereby under certain circumstances some organisms can intercalate intermediate tissue types between host tissue and graft tissue. How can that be accounted for in our scheme? If a cell's identity is represented by its clock sequence, how can a neighboring cell read that sequence, if as we have assumed, it is located on some chromosome deep in the nucleus of the cell?

## K. External Clock Readout

One possibility might be by having the clock sequence itself code for whatever polypeptide may correspond to its nucleotide sequence. This amino-acid sequence would then differ from cell to cell, and if it was positioned at the cell surface (or close enough to it), it would allow for external detection by specialized receptors in a neighboring cell. This polypeptide would then serve as an external readout of the internal clock. Cells with similar internal clocks will have similar external readouts; cells with differing internal clocks will have differing external readouts. It should be noted that it is not necessary that the readout polypeptide be a separate compound. It could just as well be an extension onto a foundation polypeptide such that in all cells of the organism the first $k$ amino acid residues of this compound are identical; after that point, additional amino acids are added according to the codons into which the internal clock sequence can be partitioned. This latter extension onto the foundation will differ from cell to cell and will be the substrate upon which cell recognition receptors can act. It will be convenient to define the state of the readout polypeptide up until residue $k$ (which could be 0, but not necessarily) as zero-state.

## L. Reading Frames

For simplicity we have not taken into account the specification of a reading frame, i.e., that the internal clock can be contiguously partitioned in three possible ways. The author's simulation program mentioned in III.C produces a series of computer generated sample clock sequences alongside of which are the corresponding amino acids for each of the three possible reading frames. This is done for both strands of the double-stranded clock DNA sequence at each step, although in general, transcription of any given locus occurs on only one strand.

Note that we are not constrained to assume that there can be only a single type of readout polypeptide, which continuously varies (changing by one residue at a time) such that on the basis of its sequence alone a cell's identity is determined. Conceivably, there could be a set of polypeptides which differ in discrete fashion, each of whose existence is governed by specific clock/target interactions, with the particular unique combination found in any individual cell serving to announce that cell's identity.

## M. Explanation for $3^{rd}$ Position Wobble in Codons

We now note an interesting observation that has been made regarding the pattern of the previously mentioned redundancy in codons. In almost all cases, the degeneracy occurs in the final member of the triplet. This phenomenon has been referred to as the third position wobble. In mathematical terms, that represents the least significant digit. A consequence of that for our scheme is that in many cases in which the internal clocks of two cells are similar, meaning that they have descended from the same branch of the developmental pathway, their external readouts will not only be similar—they will be identical. This means that the recognition receptors do not have to recognize each individual cell, but only the general cell types—a much easier task. Had the wobble been in any other position, we would have had a confusing situation in which cells that are more closely related differ to a greater degree in their external readout values than do cells that are more distantly related.

## N. Chimeras

Another point which may possibly be understood by this scheme has to do with experiments that have been done on mouse chimeras. It has been demonstrated that up until the eight cell embryonic stage, it is possible to combine the cells from two separate mice embryos (for example, a black mouse and a white mouse) and end up with a single, normally developed mouse (black and white [38]). Perhaps the importance of the number eight can be understood by this model. Before the eight cell stage (3 divisions), there are not enough base pairs in the clock to produce a single codon. Therefore, the readout polypeptide is still in zero-state.

Although, at first glance, this simplistic explanation of the developmental process in fused embryos may seem satisfying, there is a disturbing conceptual problem to be discussed shortly for which the author does not have a ready solution.

## O. Chemoaffinity and Neural Development

Another point which might be mentioned is that this scheme could possibly serve as a preliminary step in understanding the means by which neural connections are formed. Much effort has been expended in trying to understand how a nerve cell whose body may reside at a considerable distance from some other cell upon which it is supposed to synapse, is able to direct its axon to that precise location [39]. According to our scheme, this is accomplished by tracking the progression of readout polypeptides it encounters along the way in order to guide itself to the desired cell. If the progression of polypeptides seems to indicate that it is headed in the wrong direction, it can then make a course correction. Note that the scheme, as presented here, cannot completely explain the precision of synaptic connections. This is because it is currently thought that in many cases, not only is it necessary for a neuron to synapse on the correct post-synaptic cell, but it must form that synapse on a precisely determined location on that cell. The weighting of the contribution from each pre-synaptic cell on the decision making process of the post-synaptic cell (as to how to respond, e.g., whether or not to generate an action potential) is determined by the location of the synapse. The farther away the synapse is from the soma within the dendritic tree, in general, the more delayed and weakened is its contribution to the integrative process. Proper functioning of the nervous system is thought to depend on such precision wiring, emphasizing certain inputs and attenuating others. However, our approach does not tell us anything about how, within a given cell, those factors which guide incoming axons could come to be located at exactly the proper locations within the cell (assuming that such a level of precision is indeed necessary, and could not be provided by the external environment alone, i.e., using other neighboring cells as landmarks). It is worthwhile to note, though, that recent opinion is that a rough layout of the interconnections may be achieved by adhesion molecules working in a manner similar to what we have described (although the number and uniqueness of these molecules is open to debate). The further fine-tuning and delicate refinement of the individual synapses required for correct operation may be performed by various "test-firings" and other means of functional validation [40, 41].

### P. Clock Adjustment by External Factors

An important issue which still must be addressed in order to close the logical loop is how the recognition receptors can control and modify the development process. It would seem that just as the internal clock needs the ability to change the external readout to reflect internal events, for which there is a simple conceptual mechanism for accomplishing this, namely, the transcription process; there should similarly be a means for the external recognition receptors to adjust the internal clock to reflect external events such as the presence and cell types of its immediate neighbors (Fig. 5). However, a method for accomplishing this is much more difficult to conceptualize. Although certain polypeptides do interact with certain DNA sequences, as in the case of transcription factors which turn on and turn off various genes by binding to promoter and inhibitor regions of the genome; as far as is known, there is no simple systematic correspondence between the composition of a polypeptide and the specific DNA sequences to which it will bind. More work, therefore, remains to be done on this point before the model is logically complete. Mathematically, this goal would correspond to the property that in a finite state machine, the next state may depend on the inputs (the presence of a particular environment), as well as the current state (the internal state of development as reflected by the clock sequence).

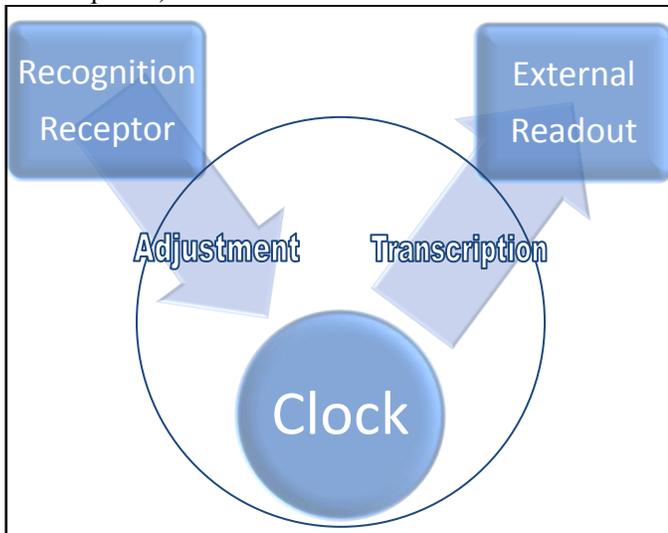

Figure 5. Schematic diagram of a single cell (circle) shows proposed bidirectional interaction between its internal clock and external environment via clock sequence transcription and recognition-receptor initiated clock adjustment. (Intermediate steps of pathways are not shown.)

### Q. Carcinogenic Agents

However, if we are willing to accept this possibility, that the recognition receptor polypeptides can indeed initiate a modification of the DNA clock sequence in a precise and systematic manner depending on the values of the readout polypeptides of neighboring cells, an interesting corollary obtains which may shed light on a possible pathway by which chemical carcinogens may act. Certain compounds may have an ability to bind with the recognition receptors of a particular cell in such a manner so as to trick the receptors into thinking that they are in the presence of neighboring cells whose presence requires a clock adjustment to be made to that particular cell (as we proposed may be necessary in the case of a graft, for example, in order to effect a smooth transition in cell types between host and graft). If this is the case, the recognition receptors will then dispatch a message through the chain of command to start altering the DNA based on this incorrect information. Since, as we have postulated, there exists machinery to do just that, dangerous results are to be expected. Interestingly, work has recently been done on developing assays to detect exposure to suspected carcinogens by looking for DNA containing cross-linked proteins in certain locations [42]. If our theory is correct, those proteins may have been deliberately sent there in error by the recognition receptors. Note that there is nothing to prevent one from simply viewing carcinogenic agents as highly toxic compounds which enter the cell and wreak all manner of havoc; effecting mutations and general pandemonium in the cell chemistry by brute force alone. However, we are suggesting that maybe there is a method to the madness. They cause an honest mistake to be made by the recognition receptors of a particular cell due to their similarity to the readout polypeptide chemistry of some other cell which either has played (in the former's developmental history) or could potentially play a role as an important neighbor. Drawing upon the processor analogy once more, our perspective is that the action of some carcinogens may be viewed as a case of garbage in, garbage out; whereas the brute force approach would view their action as a hardware crash. [Later, we will distinguish between mutagenic and non-mutagenic carcinogens. Mutagenic carcinogens may work via the toxicity approach, while non-mutagenic carcinogens could work via the receptor approach.]

Another phenomenon which is accounted for by this scheme is the existence of teratomas, which are undifferentiated growths that can be produced by grafting a cell from a developing embryo onto a differentiated cell such as a kidney [43]. This follows from the fact that the recognition receptors will be thrown way off by the sudden abrupt appearance of readout polypeptides belonging to a cell from such a different time and place in developmental history. In order to try to achieve continuity, which is one of their main tasks, they may attempt to make major changes in the DNA clock sequence which will completely disorder the current state. The receptors may not realize that continuity is actually not desirable or even achievable in this case, but nevertheless, still proceed with that intention, causing chaos to ensue. Conceptually, the situation is very similar to that which we proposed above to explain the action of some chemical carcinogens.

The main problem with this explanation is that it would seem to actually over-predict the incidence of such occurrences, since many types of grafts seem to take hold without any problems, and only under certain unusual conditions does this type of wild growth occur. One is forced to make some type of distinction such as that if two cells have developed along a common path (i.e., share a common clock sequence up to a certain point), then only minor internal changes are necessary to achieve compatibility; whereas if the paths are highly divergent or differ greatly in length, then major, perhaps catastrophic changes are indicated.

## IV. NECESSARY EVIDENCE

This concludes our development of the microcode model. Evidence needed to support this hypothesis would need to be found in terms of differences, even slight, in the DNA of different cells from the same organism. In addition, external membrane chemistry would need to reveal a systematic variation or progression among different cells in the amino acid sequence of some protein or proteins. A sufficient set of molecules (receptors) that are able to detect these differences would also need to be found, so that the differences in the readout polypeptides could be leveraged. One should not rule out the possibility that the readout polypeptides themselves perform a double duty, and are also able to serve as receptors for the readout polypeptides of other cells.

## V. POTENTIAL PROBLEMS WITH MODEL

Undoubtedly, the most difficult challenge is to satisfactorily explain those experiments which seem to demonstrate that the DNA of all cells in a given organism is identical (or at least does not undergo irreversible change), such as the fact that it is possible to transplant a nucleus taken from certain cell types of a fully differentiated adult frog into an enucleated frog egg, and still produce a normal tadpole [44, 45]; and more recently, the achievement of cloning in mammals (sheep) using genetic material from mammary gland cells to produce viable offspring [46]. In other words, cells are potentially totipotent (can be induced into playing any role) up to an advanced stage of their development. One is forced to take the stance that the existing polypeptide machinery in the remainder of the egg cell is able to properly reset the clock of the differentiated cell's DNA, if all subsequent events are reversible. Despite the above difficulties, some level of DNA rearrangements may not necessarily be out of the question, and are known to occur in the immune system. (Even conventional theories which do not postulate any differences in DNA sequences from cell to cell are faced with some conceptual difficulties from these experiments, such as how all the bound transcription factors which control gene expression and define the developmental state of a given cell are suddenly dislodged and/or disabled and replaced with a completely different set which effectively reverses and resets the cell's entire developmental history.)

Other evidence which supports the unlikelihood of intercellular DNA differences includes reassociation experiments using DNA from different organs [47], and various comparisons of chromosome morphology. However, neither of these may be fine enough to rule out the small-scale variations which are all that is necessary to operate a clock.

Another possible area of difficulty for this scheme is one we alluded to in our discussion of fused embryos. The problem is this: Consider a case of two identical cells (containing identical clock values and identical readout polypeptides) from two identical embryos. If they are now combined into one embryo, how do they coordinate their clocks so that each cell proceeds along a different branch of the developmental tree, rather than duplicating one branch and omitting the other? There must be some negotiation process (or in digital design parlance, a handshake) between them so that one agrees to switch its role over to the other branch. That, in and of itself, is not terrible, as we have proposed that internal clocks can be adjusted in response to the external environment. However, since the two cells are initially identical in all respects except for the fact that they stem from different embryos, what arbitrating factor selects one cell over the other to be the one to go through the tiresome ordeal of a clock adjustment, while its companion just sits and relaxes? The analogous situation in digital design would lead to a metastable state, where a device sits in limbo because it can't make up its mind as to which of two stable states it will settle into.

Further quantitative analysis needs to be done in establishing a metric and error threshold for clock/target interactions, since a given cell's clock sequence will often match with a subset of some target sequence not intended for use by that cell or vice versa; and one might reasonably expect that if the similarities were close enough, that may be sufficient to falsely trigger the subsequent cascade of events that occurs upon achievement of a correct match. However, to reduce this probability would require longer clock and target sequences, and hence, a more complex nucleotide selection rule. Improved synchronization could possibly be achieved by the use of unique flag sequences which cannot be generated by the normal selection rule (such as, for example, a string of repeated nucleotides), to designate the beginning and/or end of a clock or target region. Such a device is routinely used in digital communications protocols to avoid the need for a fixed frame size [48].

## VI. CONCLUSION

The general idea of modeling the cells of an organism as an array of automata operating in parallel and exchanging information with each other is not new[49]. This paper merely attempts to show a general route painted in the broadest of strokes, by which that can be accomplished using some known properties of cells. With the recent demonstration by Adleman [50] that molecular computing using DNA can be used to solve certain simple problems in the lab, some of the ideas discussed here should sound a trifle more plausible. (Note that the sequences used in that experiment were static, i.e., were all present *a priori*, and were selectively joined via the use of specific complementary splints anchored through base pairing interactions. Our mechanism, however, requires a dynamic clock sequence that is able to advance without the use of templates.)

As things currently stand, technology may not yet exist to directly verify the existence of the types of sequence variations (in both cell DNA and cell or membrane polypeptides) upon which this scheme is based. For the present, evaluation may have to be limited to comparison of the known body of knowledge against the ideas presented here, and examination of future results in light of these ideas.

## VII. POSTSCRIPT

This concludes what is practically a verbatim presentation of the 1997 paper submitted to Nature, save for a few grammatical corrections, and the addition and improvement of some of the figures. Looking back at progress in the field during the past 16 years since the paper was written, we now present additional arguments to support the case that this model should be considered as part of the scientific discourse. We cannot directly argue the correctness of the model, since that will require additional very specific evidence that may not yet be available, as we will outline. Nevertheless, we believe that the model should be judged on its coherence and consistency, and on its ability to explain a number of important issues in the areas of molecular biology and cancer research. A scientific theory may also be judged on the basis of the predictions it makes, and our original paper made some clear predictions which appear to have been confirmed, as we noted in the Preface. The most important was the idea that noncoding DNA serves as the logic upon which gene regulation is based, and is not "junk DNA".

Before listing these various arguments and advantages, we provide a brief summary of the paper.

### A. Summary

The Microcode Model proposes that at every cell division, there exists an active, asymmetrical, single-nucleotide-polymorphism (SNP) generator that either adds or modifies a single DNA base-pair in each cell, such that the DNA molecules of the two daughter cells end up differing from each other in a single position. In this article, we chose to work with the additive route, for convenience. We showed that because DNA has 4 possible bases, that this asymmetric outcome is possible with a simple, predefined, ordered scheme. Over the course of multiple divisions, there will be at least one region in the DNA of every cell that will differ by a few base pairs from that of all other cells in the organism. (We calculated that, in principle, approximately 27 base pairs are sufficient to tag each cell in a human with a unique identifier. This would be almost negligible compared to the total 3 billion base-pairs in the human genome.) We termed this a clock region.

We then proposed that there exist other regions shared by the DNA of every cell, which serve as targets, and may be situated near particular genes which are supposed to be turned on in certain specific cell types. But like a lock and key, or like a packet designed for a specific IP address, the targets will only interact within those cells that have specific clock values.

These interactions consist of hybridizing a clock sequence with a matching target sequence. The result of these interactions will be selective looping, recombination, translocation or other events that differ from cell to cell, because of the differing clock sequences that all cells have. In cell A, the clock sequence may loop with a target nearby, while cell B's clock does not match that target, but does match a target farther away. The net result is that the topology of DNA will then vary from cell type to cell type. (We believe that this prediction has also been confirmed in the past 16 years but its significance is again, not understood.) The importance of this is, in fact, that specific transcription factors may be able to bind to DNA in one configuration, but not in another. Therefore, there will be differential expression of genes in one cell type as compared to another.

We proposed that errors in this process can cause the translocations and chromosomal rearrangements found in cancer. If a mutation occurs in the clock region which causes it to match with some non-intended target located at an inopportune location, steric or angle strains may then cause the chromosome to break, as it was not designed to be stretched that far, or bent that sharply. Thermodynamically, perhaps the energy of formation of multiple hydrogen bonds (in clock/target base-pairing of a certain minimum length) exceeds the bond energy of some particular covalent bond in the sugar-phosphate backbone, and thus is strong enough to cause breakage of the DNA molecule.

The second half of the paper discusses how perhaps mechanisms exist to deliberately modify the clock sequence as necessary, in order to respond to changes in a cell's environment. For example, if certain cells are destroyed by trauma, one may need to reset the clock value of neighboring cells to regenerate the missing types; or if a cell is grafted, the neighboring cells may need to read the newcomer's clock to determine how to smoothly integrate with its cell type. For this reason we have used the word 'Dynamic' in the title. We believe DNA may be written to, as well as read from.

### B. List of Key Strengths

We now collect in concise form what we believe are the main advantages of the Microcode Model, some of which we have alluded to earlier, but which will be further emphasized here, and some of which we were not aware of at the time of submission of the original paper, 16 years ago.

#### 1) 4-base and complementary arrangement of DNA

The model explains very clearly why the structure of DNA is complementary; and why coding is quaternary (using four bases), and not binary (using 2 bases). As we suggest in the paper, this allows for asymmetric SNP generation between daughter cells to occur which is necessary for our scheme to work.

#### 2) Noncoding DNA

Our model well-accounts for the role of noncoding DNA, and was proposed 16 years before the results of the ENCODE project were made public. Noncoding DNA contains the clock and target sequences that are the directors of gene expression. It also forms the precise shapes that are needed to assist or hinder specific transcription factors from binding. The lengths and sequences between genes are critical.

### 3) DNA Looping

The model explains the significance of DNA looping, which, as in the Preface, is not considered well-understood. Our model suggests that DNA looping will differ from cell type to cell type depending on the clock, and this is crucial for differentially activating particular genes, as in Fig. 4.

### 4) DNA-DNA Searching

The sequence selectivity necessary for gene regulatory networks is borne by DNA-DNA interactions themselves, which are the most suited for that purpose, because of the natural base pairing properties of DNA. I.e., DNA is best able to search for a complementary sequence with which it can interact, and can do a better job at searching for a specific DNA sequence, than can proteins such as transcription factors. Why not build a system in which DNA can leverage its own superb capabilities? Adleman experimentally verified this search capability in his work on solving mathematical problems via DNA, and further confirmed our way of thinking, that one molecule of DNA can search for matching patterns in other molecules of DNA. However, Adleman worked with single-stranded DNA, while our model requires similar behavior in double-stranded DNA. Nevertheless, mounting evidence shows that this is plausible, as we discuss in VII.B.9)

### 5) Generalized Transcription Factors

Our scheme takes a heavy workload off the shoulders of the transcription factors (TF's). Rather than having each TF recognize a single DNA sequence, which requires numerous custom-built TF's; instead, it allows TF's to recognize specific topologies which may occur at different locations in different cells, but are essentially similar. In cells of type $A$ where DNA is looped in one particular configuration, the TF fits at point $X$, and initiates transcription from there. In these cells, at location $Y$, where topology is straight, the TF does not bind. But in another cell of type $B$ with a different topology, the same TF's might find a proper fitting loop next to a different gene than in the first cell, say at location $Y$; and will transcribe that gene, instead of the gene located at point $X$ which was transcribed in the first cell. This in effect allows for modularity in the design of TF's, which can now reuse the transcription engine, but effectively operate on different genes in different cells. In other words, use a different plug to fit the electrical outlets of different countries, but keep the same electronics of the device.

### 6) Computational Power of DNA

This scheme makes use of the full computational potential of DNA. It is able to search and distinguish down to the single base level among different sequences, and to act as a dynamic platform, in terms of modifying bases, when necessary, to accommodate changes in environment. All existing models do not harness the computational potential that DNA is naturally endowed with to actually act as a full-fledged processor, and not as a mere template or blueprint for proteins.

### 7) Precise specification of properties of cells along particular developmental branches

The model enables cells which share common lineage up until a certain point to use similar developmental controls in the form of the first few base-pairs in their clock region which are common to both. Using a short target will allow the DNA of all these cells to undergo the same interaction. At the point where the cells must diverge in function, using a longer target incorporating the remaining base pairs of the clock will selectively place these cells on a different developmental path. Think of it as an area code for a phone number. If a message is intended for everybody in the area code, then a phone company can choose to look at the first digits of the number, and ignore the rest. However, if only certain recipients (for example, a single exchange) are to receive the message, then one can look at further digits to separate the two groups of people. This can be done on a digit by digit basis with this scheme, using as few or as many digits as are needed to specify the groups of cell types which should respond. This allows for great precision and efficiency, and for sharing the same machinery in all recipients.

### 8) G-quadruplexes

There is a rapidly growing literature on quadruplex DNA [51-53] which on the surface seems quite relevant for our work. These are 4-stranded structures often formed of 2 double-stranded DNA regions that have come together in a stable configuration, often in some kind of loop. They have been found with great frequency near promoters [54], and have also been tied to telomeres in various reports, with some researchers linking them to cancer [55]. They seem to require guanine-rich sequences to form. We note this as a curiosity, as our work depends on formation of DNA-DNA loop structures, and we wonder whether they might be related to the clock-target interactions we have envisioned.

### 9) 3-D DNA Conformation

Lieberman and colleagues [56] cited in [57] have recently developed a method for freezing the 3-D structure of DNA in time by means of a chemical treatment, and then chopping it up with restriction enzymes such that pieces of DNA that were in proximity remain together and can be identified. They have found that certain genes are not activated unless in contact with distant DNA sequences that may even lie on different chromosomes. This appears to correspond with our idea that DNA-DNA interactions are important for regulation. This idea is gaining currency in general [58]. Perhaps different cell clocks will produce different 3-D conformations in different cells, and thus will turn on and turn off different sets of genes as appropriate for each tissue type. To our knowledge, the explicit formulation of a driving force behind this differential activity has not been definitively articulated by any workers in the field. That is the primary contribution we are attempting to make. It all starts with a single, systematic, DNA base-pair change, and everything follows from there.

### 10) Anomalies suggest a new viewpoint.

Rather than categorizing the growing number of irregularities, such as quadruplexes and DNA-DNA interactions as anomalies, it would seem there is a limit to what can be accounted for in the old model. The main drawback of the conventional model is that it treats DNA as a static molecule, identical in all cell types, which has no real role other than as a cookbook of recipes for proteins. That is why the staggering amount of noncoding DNA was such a puzzle for so long. It is also why the number of actual genes in human is so surprisingly small, about 20,000--less than in many simpler organisms. It is also partially the reason why in the past 10 years since the human genome was sequenced, not a single disease has yet been cured, a fact which led to a rather contentious exchange reported in a recent New York Times article [59]. It is not the genes themselves that are the mystery, but the logic that governs precise expression at the right time in the right cells, and the activity and 3-dimensional structure of the gene products (polypeptides). It is our belief that the older viewpoint must give way to new thinking about DNA, and we have demonstrated that DNA has the capability to function in far more sophisticated ways. At this point, it should be more surprising if this capability were unused and wasted, than if the elements of our model were actually correct.

To belabor this point once more, why build complex, custom-designed protein machinery, with numerous intricate components to regulate each individual gene, when simple generic DNA-DNA interactions can do much of the same work? (Occam's razor.)

### 11) Single Nucleotide Polymorphisms (SNP's)

The model fully explains the significance and potentially major regulatory changes that a single SNP, even in a noncoding region, may have on a cell's development. A clock or target value which is altered by even a single base may completely change the expression profile of the entire cell as in Fig. 4. Just like the change of a single bit in the operand of an `IF` statement in a high level computer language can completely change the execution of the program, so can a change of a single base in the DNA of a cell. It follows that two individuals who differ in only a single non-coding base-pair may develop in completely different ways, e.g., different heights, etc.

### 12) Chromosomal translocations of cancer

The model accounts in a straightforward way for the physical and often predictable chromosome breakages which occur in particular types of cancer. Current models which look at altered levels of expression or composition of certain proteins, or look for mutations in particular genes simply cannot explain why a chromosome should snap or rearrange. Our model can, as outlined above in section III.I and in the summary, VII.A.

### 13) Gene over-expression in cancer

In other manifestations of cancer, a gene may be over-expressed, because TF's are incorrectly signaling a need for transcription. This model accounts for this behavior as well, for it links the binding of TF's with the local topology of DNA, which is in turn determined by clock/target interactions. An incorrect clock or target value may alter topology in a way that enables an errant TF to gain access.

### 14) RAM vs. ROM

We stress that there are two components to the computational model we assume for DNA. The first is the ability to execute search functions, and to guide the development according to a predefined logic which is manifest in the software that is represented by the non-coding DNA sequences. For that reason, we call the model a Microcode Model. However, microcode can be hardwired, as in Read Only Memory (ROM) which allows a device's program logic to be fully executed, but cannot be altered. Here we have taken a second step, as well, and propose that not only is noncoding DNA functioning as microcode, but the program logic can be altered on the fly, as necessary, such as in the case of unexpected events like trauma or grafts, etc. For this reason, we add to the title the phrase Dynamic DNA Processing, which accents the fact that we believe DNA can function as Random Access Memory (RAM), i.e., read and write memory. We postulate that not all mutations are harmful and deleterious, but that mechanisms exist to induce mutations in order to change a clock, when necessary. We opined that perhaps, certain carcinogens which are not mutagens operate in this manner (and they exist [60]), by fooling the cellular machinery into thinking an event has occurred which requires a clock adjustment. They may be chemical analogs of the active sites on membrane proteins that might serve as readout polypeptides, as described earlier.

### 15) Testability

The model is completely testable and falsifiable. It depends crucially on the existence of single- or multiple-nucleotide polymorphisms between the DNA of different cells or cell types within the same organism (and within the same individual). A method by which these differences are both generated and harnessed is the central contribution of this paper. Sequencing studies which are capable of testing this are expected soon, but were not practical when this paper was first submitted 16 years ago. At that time, sequencing of even a single human genome had not been completed. It is our hope that this work will serve as motivation for carrying out direct comparisons not only of the genomes of different individuals within a species, but of different cell types within an individual. This seems to be the only conclusive evidence that can actually prove or disprove the truth of this model. The other points we have stated could quite possibly be nicely explained on the basis of the model, but do not seem to offer concrete proof of its validity, as perhaps other alternative mechanisms are in play.

### 16) Healthy clones from cancerous cells

There have been studies which have demonstrated the ability to clone a healthy animal from a cancerous cell [61]. If, as is conventionally thought, cancer is a defect in particular genes, then how can the offspring develop normally, if there are defects in the genetic material of the donor cell? The same gene that caused the disease in the parent should cause the disease in the offspring. This seems to suggest that perhaps cancer is not a defect in any gene per se, but rather in the general topology of the DNA, which may be governed by clock/target interactions, and can be reset upon insertion into an egg.

Conventional models, on the other hand would attribute this to possible epigenetic factors (DNA methylation, and other non-sequence-related changes in the chemical environment) to explain this phenomenon.

### 17) Alternative RNA Splicing

This model can account for alternative splicing [62, 63] as a method for creating multiple types of polypeptides from a single gene. When the clock of a particular cell reads one value, it may cause DNA/RNA interactions that lead to one form of splicing. In a different cell type, with a different clock value, different DNA/RNA interactions may occur that lead to a different splicing pattern, such that different polypeptides are produced in each cell. In other words, in one cell, certain exons or introns may match the clock value and be preserved, while in a second cell with a different clock value, they may not match and be excised, or vice versa. Perhaps the RNA transcription of noncoding regions reported by the ENCODE consortium may play a role in controlling or administering this type of alternative splicing or translation scheme.

### 18) Gene therapy initiates tumorigenesis

Gene therapy continues to be a vexing problem, and often the injection of a correct copy of a gene into a patient who has a defective copy of that gene ends up causing tumors, even when the gene in question is completely unrelated to cancer. Conventional models have difficulty accounting for this. Our model may explain what is happening. The insertion of a gene into a random place in the genetic material may disrupt the correct topology of the DNA, say, by moving a given target farther away from the clock than it should normally be. This can cause stretching strain in a clock/target interaction, and could potentially break a chromosome, or disrupt proper regulatory proteins from binding due to the abnormal topology. Our model provides guidance in how to avoid this problem. The incorrect gene needs to be excised, and the correct gene needs to be inserted at precisely the correct location. This is not done currently, but rather a promoter is attached to the gene, and it is randomly injected. While it may indeed be expressed, it may potentially cause damage for the reasons we explained. The model also predicts that even if no tumors are generated, the chances of expression may themselves be low, as correct topology for transcription may not be maintained.

### 19) Reprogramming

The model may provide insight as to the nature of reprogramming a cell from one type to another, as in cloning or stem cell research (particularly, induced pluripotent stem cells or IPSC's). Perhaps, part of the reprogramming task consists of altering the clock from one value to another which will change the future developmental behavior of that cell and its descendants. When we first developed the model 16 years ago, we were disturbed by the fact that it seems to allow for cloning and reprogramming, which had never been successful at that time. Then, right around the time we submitted the paper, first reports of cloning were emerging which we mentioned in the paper. We had not time to digest whether they supported or contradicted the ideas expressed in our model. However, in the ensuing years since the work was published, some evidence has emerged that many cloned animals suffer from cancer and other developmental defects at an increased rate compared to the general population, and that many of these embryos do not develop normally from the outset [64]. Dolly, the first clone, died at age 6.5 from lung cancer, earlier than the normal 12-year life expectancy for sheep [65]. This might be explained based on our clock/target model as resulting from the fact that the clock value in the DNA of the somatic cell from which it was taken is not equal to that in a normal egg cell. This may cause mismatches at some stage. However, as above, it must still be explained why the cell is able to develop normally until that point. Some researchers believe that the abnormally short telomeres which are often seen in clones may be the culprit that causes this early death by artificially advancing a cell's age. In our model one might postulate that the telomeric region may play some role in the functioning of the clock/target system.

On the other side of the coin, the FDA is decidedly positive about the quality of cloned animals and their potential for use as food [66] based on a thorough review of the evidence through 2009.

Clearly this issue requires more work to resolve. We only mention it, because the mechanism by which reprogramming works, for better or for worse is relevant to our model. At this time, we can say only that perhaps research on sequencing and comparing the genomes of reprogrammed cells may offer valuable insight.

Regarding IPSC's there are two main methods by which they can be produced. Transfection with genes that control transcription, or by molecular mimicry using external drug-like compounds [67] cited in [68]. The first has risks of cancer, as expected, based on reasoning we discussed regarding gene therapy. The second demonstrates that external factors can reverse state of development. Our model would mandate that this actually causes a write operation to the DNA clock, as before. Conventional thought would explain as alteration of bound transcription factor array. The only way to distinguish between these two competing explanations is with accurate sequencing, as in the next section.

### 20) Accuracy of current sequencing methods

Finally, we acknowledge that the central dogma of molecular biology continues to be that DNA is transmitted identically from parent to daughter cells, and that any variations in this process are to be considered errors or mutations which must be corrected by special error-correcting machinery, or are liable to cause great problems. This has made our model a very contentious proposition. However, we ask whether there is clear enough evidence at this time to rule out a single nucleotide change in the replication process. Is genomic sequencing mature enough at this stage to detect every SNP? Until recently, at least, many assumptions were made in the sequencing process in terms of matching and ordering fragments of DNA to reconstruct a full chromosome. When the emphasis was in looking for genes, one could afford to be lax in positively identifying every single base pair, especially in the noncoding regions. 99% or 99.99% accuracy was sufficient, etc. But to evaluate a model such as we have proposed, it is imperative that we sequence down to the single base-pair level, in order to compare different cell types base for base. For example, an error rate of only 1 in $10^4$ (99.99% accuracy) still allows for $10^5$ errors per human genome which would totally miss the fine variations we seek.

*21) DNA variations among different tissue types*

To our surprise, a recent NY Times article [69] reported on the work of Alexander Urban and others [70] which described a number of types of observed variations among the DNA of different tissues within the same individual. This point had been the most elusive and difficult for us to prove, and probably the most controversial part of our proposition in the past 16 years. It is also the most crucial piece of evidence that we sought, and perhaps indicates that a scheme such as we have described can indeed be plausible. The previous absence of this data may be the reason this paper was consistently rejected by reviewers after being submitted on at least seven separate occasions throughout the years to well-known journals, including Science, Nature, Nature Genetics, Cell, Journal of Theoretical Biology, Journal of Biomedical Hypotheses, and most recently to IEEE Transactions on Biomedical Circuits and Systems. We also acknowledge that this is a theoretical paper and does not present any new data, which is unusual in the world of biology, and this could also hinder its acceptance. Nevertheless, we feel it has a place in the scientific literature.

## VIII. Final Remarks

We believe we have demonstrated that the potential exists for sophisticated DNA processing based on existing hardware in the cell. No different than the 1's and 0's of digital computers. But this requires an asymmetry (clock) to initiate the process. Whether such actually occurs, we can't yet tell. Nevertheless, perhaps it is worth pursuing.

Because many predictions made in our original paper have since been found to be accurate, we would welcome further discussion and opportunities to collaborate with individuals or groups who may be interested in further developing this approach.

## IX. Acknowledgment

The author wishes to thank Professor David Gifford of MIT for reviewing an early version of the manuscript, and for providing helpful advice and some of the references in the area of Computational Molecular Biology. Leonid Litvak made many incisive comments. Ben Rothke proofread. Special thanks to Gert Cauwenberghs for his encouragement and faith in this work, and for his painstaking guidance throughout my thesis research (which was actually in the unrelated field of auditory signal processing).

I will always owe an immense debt of gratitude for the kindness of the administrators and faculty of the Harvard-MIT Division of Health Sciences and Technology for accepting me and providing a superb graduate school education within the outstanding facilities and intellectual climate of these two venerable institutions of higher learning. Special thanks to the Program Directors: Nelson Kiang, Lou Braida, Bertrand Delgutte and Bill Peake, and to my thesis co-advisor, Tom Quatieri. This work was supported in part by N.I.H. training grant 5T32 DC00038.

I would like to dedicate this work to the memory of my father, Dr. Myron Jacobson, a thoracic surgeon who coauthored with Paul Lauterbur some of the early papers on MRI [71] and cited in [72] who recently passed away at the age of 80, to my mother, Dr. Janice Jacobson Sokolovsky, and to my wife, Shira Jacobson, for all their help.

```
Generation 10:
  Cell   1:
    Base Pair:  1 2 3 4 5 6 7 8 9 10    Frame:  1        2        3
               --A--C--G--T--A--C--G--T--A--C--         TYV      RTY      VR
                 |  |  |  |  |  |  |  |  |  |
               --T--G--C--A--T--G--C--A--T--G--         CMH      ACM      HA
  Cell   2:
    Base Pair:  1 2 3 4 5 6 7 8 9 10    Frame:  1        2        3
               --A--C--G--T--A--C--G--T--A--T--         TYV      RTY      VR
                 |  |  |  |  |  |  |  |  |  |
               --T--G--C--A--T--G--C--A--T--A--         CMH      ACI      HA
  Cell   3:
    Base Pair:  1 2 3 4 5 6 7 8 9 10    Frame:  1        2        3
               --A--C--G--T--A--C--G--T--G--T--         TYV      RTC      VR
                 |  |  |  |  |  |  |  |  |  |
               --T--G--C--A--T--G--C--A--C--A--         CMH      ACT      HA
  Cell   4:
    Base Pair:  1 2 3 4 5 6 7 8 9 10    Frame:  1        2        3
               --A--C--G--T--A--C--G--T--G--C--         TYV      RTC      VR
                 |  |  |  |  |  |  |  |  |  |
               --T--G--C--A--T--G--C--A--C--G--         CMH      ACT      HA
  Cell   5:
    Base Pair:  1 2 3 4 5 6 7 8 9 10    Frame:  1        2        3
               --A--C--G--T--A--C--G--C--G--T--         TYA      RTR      VR
                 |  |  |  |  |  |  |  |  |  |
               --T--G--C--A--T--G--C--G--C--A--         CMR      ACA      HA
  Cell   6:
    Base Pair:  1 2 3 4 5 6 7 8 9 10    Frame:  1        2        3
               --A--C--G--T--A--C--G--C--G--C--         TYA      RTR      VR
                 |  |  |  |  |  |  |  |  |  |
               --T--G--C--A--T--G--C--G--C--G--         CMR      ACA      HA
  Cell   7:
    Base Pair:  1 2 3 4 5 6 7 8 9 10    Frame:  1        2        3
               --A--C--G--T--A--C--G--C--A--C--         TYA      RTH      VR
                 |  |  |  |  |  |  |  |  |  |
               --T--G--C--A--T--G--C--G--T--G--         CMR      ACV      HA
  Cell   8:
    Base Pair:  1 2 3 4 5 6 7 8 9 10    Frame:  1        2        3
               --A--C--G--T--A--C--G--C--A--T--         TYA      RTH      VR
                 |  |  |  |  |  |  |  |  |  |
               --T--G--C--A--T--G--C--G--T--A--         CMR      ACV      HA
  Cell   9:
    Base Pair:  1 2 3 4 5 6 7 8 9 10    Frame:  1        2        3
               --A--C--G--T--A--C--A--C--G--T--         TYT      RTR      VH
                 |  |  |  |  |  |  |  |  |  |
               --T--G--C--A--T--G--T--G--C--A--         CMC      ACA      HV
  Cell  10:
    Base Pair:  1 2 3 4 5 6 7 8 9 10    Frame:  1        2        3
               --A--C--G--T--A--C--A--C--G--C--         TYT      RTR      VH
                 |  |  |  |  |  |  |  |  |  |
               --T--G--C--A--T--G--T--G--C--G--         CMC      ACA      HV
```

```
Cell 11:
    Base Pair:  1  2  3  4  5  6  7  8  9 10     Frame:  1        2        3
               --A--C--G--T--A--C--A--C--A--C--           TYT      RTH      VH
                 |  |  |  |  |  |  |  |  |  |
               --T--G--C--A--T--G--T--G--T--G--           CMC      ACV      HV
Cell 12:
    Base Pair:  1  2  3  4  5  6  7  8  9 10     Frame:  1        2        3
               --A--C--G--T--A--C--A--C--A--T--           TYT      RTH      VH
                 |  |  |  |  |  |  |  |  |  |
               --T--G--C--A--T--G--T--G--T--A--           CMC      ACV      HV
```

**Figure 6.** Output of computer simulation to generate DNA clock sequences for all cells of a generation. First 12 cells of generation 10 are shown. Next to each sequence is the corresponding amino acid sequence which would be read (using standard single-letter abbreviations). Both the upper and lower strands are translated, and the amino acid sequence is shown for each of the three possible reading frames. Note that the last frame can only complete 2 amino acids, since the 10th generation produces a clock sequence of 10 base-pairs, which when shifted by 2 positions for the 3rd reading frame yields 8 base-pairs, enough for 2 codons of 3 base-pairs each, and a remainder of 2 base-pairs. Further note that although each cell has a unique DNA clock sequence, because of the 3rd base-pair interchangeability in the genetic code, however, the amino acid sequences for certain cells and reading frames are identical. As discussed in text, this may allow different cells of same tissue-types to share similar ID values for purpose of identification by other cells.

|  | Microcode Model | Conventional View |
|---|---|---|
| **Central Dogma:** | DNA varies slightly from cell to cell in a given organism. This variation of clock activates different target regions throughout the genome via selective base-pairing in a lock/key model, leading to differences in development and gene expression from cell to cell.<br><br>Cell can write to its clock as appropriate to adjust to presence of specific neighboring cell types in external environment.<br><br>Particular genes may reflect coding of clock, and hence possess variable sequences from cell to cell. These will lead to variable polypeptide products that may serve to uniquely identify each cell to its neighbors. | DNA in all cells is identical, other than through random mutations which must be prevented and corrected to the greatest degree possible, with the one exception being the immune system.<br><br>Cells do not purposely alter their own DNA.<br><br>Genome is generally constant from cell to cell with certain exceptions such as in the immune system, where rearrangements can occur. |
| **Quaternary Coding (4 Bases):** | Minimum number necessary for required asymmetries of daughter strands. | No opinion. |
| **Complementary Strands:** | Necessary for asymmetry of daughter strands. | No opinion. Happens to facilitate correct bond lengths and correct number of hydrogen bonds, either 2 or 3 depending on base-pair. |
| **Non-coding DNA:** | Necessary for clock and target regulatory regions. | Junk DNA. Possible structural role. |
| **Gene Regulation in Development:** | DNA-DNA interactions mediate binding of transcription factors (TF's) by effecting conformational changes in DNA that selectively assist or block access to TF's in addition to conventional view. | Via selective binding of transcription factors to particular regulatory DNA sequences. Types of TF's differ from cell to cell. |
| **Neural Migration in Development:** | In addition to conventional view, via numerical value of readout (surface) polypeptides which serve as individual ID number for each cell. | Via gradients in neurotrophic growth factors, and other unknown mechanisms. |
| **Short Repeating Sequences in Non-coding DNA:** | Possible function in clock/target interactions. | No opinion. |
| **Transposons:** | Communication among chromosomes regarding value of clock. | Curiosity. |
| **Limb Intercalation Experiments:** | Recognition of numerical values of readout (surface) polypeptides causes cell to alter DNA clock to appropriate value for missing limb cell types. | Appropriate expression of genes for missing cell types via other unspecified mechanisms. |
| **Cancer:** | Root-Cause: Mismatch between clock and target DNA sequences. | Multiple factors, but root not understood. |
| **Chromosomal Breakage in Cancer, Fusion Genes:** | Improper clock or target value sets off conformational changes that stress chromosomes to breaking point. | Not understood. |
| **Non-random Chromosomal Translocations and Deletions in Cancer:** | Improper clock or target value drives an abnormal recombination event. | Not understood. |
| **Teratomas (tumors characterized by mismatched tissue types):** | Attempt to write to clock to intercalate intermediate (nonsense) tissues results in severe clock-target mismatch. | Environmental signals play some unspecified role in cancer. |
| **Carcinogenic Agents:** | Directly and indirectly mutate DNA via the cell's own writing mechanisms. Appear to cell surface receptors as graft of dissimilar tissue (molecular mimics or analogs). Cell tries to intercalate, as with teratomas. | Directly mutate DNA. |
| **Success of Cloned Organisms:** | Internuclear transfer resets numerical value of DNA clock in addition to conventional view. | Internuclear transfer modifies environment and swaps bound TF's thereby simulating early stage of development. |
| **High Incidence of Cancer in Cloned Organisms:** | Incomplete resetting of clock causes mismatches with various targets. | General developmental errors. |
| **Problems of Gene Therapy** | Injection of gene into random location may cause incorrect DNA conformation, leading to breakage or low levels of transcription. | Not systematically understood, with some cases succeeding, and some cases causing severe harm to patient. |
| **Induced Pluripotent Stem Cells** | Alter clock (which can also cause tumors, as above). Secondarily to altered clock, binding of Transcription Factors is altered as in Fig. 4. | Alter complement of Transcription Factors by unknown mechanisms. |

Table 1. A short summary comparing the elements of the Microcode Model to the conventional view.